\documentclass[twocolumn,prb,amsmath,amssymb,amsfonts,superscriptaddress,floatfix,showpacs]{revtex4}

\usepackage{epsf}
\usepackage{graphicx}
\usepackage{dcolumn}
\usepackage{bm}
\usepackage{rotating} 

\begin{document}

\title{Effect of electronic interactions on the persistent current in one-dimensional disordered rings}

\date{\today}
\author{E. Gambetti}
\affiliation{Institut de Physique et de Chimie des Mat\'eriaux de
Strasbourg (IPCMS), UMR 7504 du CNRS, 23 rue du Loess, 67037 Strasbourg, France, EU}

\begin{abstract}
The persistent current is here studied in one-dimensional
disordered rings that contain interacting electrons. We used the
density matrix renormalization group algorithms in order to
compute the stiffness, a measure that gives the magnitude of the
persistent currents as a function of the boundary conditions for
different sets of both interaction and disorder characteristics.
In contrast to its non-interacting value, an increase in the
stiffness parameter was observed for systems at and off
half-filling for weak interactions and non-zero disorders. Within
the strong interaction limit, the decrease in stiffness depends on
the filling and an analytical approach is developed to recover the
observed behaviors. This is required in order to understand its
mechanisms. Finally, the study of the localization length confirms
the enhancement of the persistent current for moderate
interactions when disorders are present at half-filling. Our
results reveal two different regimes, one for weak and one for
strong interactions at and off half-filling.
\end{abstract}


\maketitle

\section{\label{sec:intro} Introduction}

Mesoscopic rings are known to carry an equilibrium current called
persistent currents when they are threaded through a magnetic flux
\cite{buttiker83}. The experimental discovery
\cite{levy90,chandra91,mailly93} of persistent currents has
motivated a number of theoretical studies. For example,
non-interacting electron theories
\cite{oppen91,altshuler91,schmid91} describe the general features
but they strongly underestimate the current present within the
diffusive regime \cite{jariwala01}. The fact of taking into
account either electron-electron interactions \cite{loss92} or
simply the degree of disorder \cite{montambaux90} leads to an
underestimation and hence, only a partial description of the
magnitude of the currents. A possible way to improve the
characterization of the persistent currents is to take into
account both disorder and electronic interactions \cite{pctheory}.
Unfortunately, this has proved to be a difficult problem to
tackle. Diagrammatic perturbative expansion in three dimension
suggests an increase in the current \cite{ambegoakar90} but does
not provide the means to conclude firmly. Therefore, we restricted
the problem here to one dimension. Surprisingly, the role of
electronic interactions is still a controversial question and some
have even suggested that such interactions may lead to a decrease
in the magnitude of the persistent currents
\cite{georges94,kato94}. On the one hand, experiments using exact
diagonalization for systems containing very few lattice sites
\cite{berkovitz93} and the self-consistent Hartree-Fock
approximation \cite{kato94,Cohen98} show that repulsive
interactions suppress the persistent currents whereas calculations
show no dramatic increase in the magnitude of the persistent
currents \cite{berkovitz93,abraham93}. On the other hand, DMRG
numerical studies carried out on spinless fermions claim to obtain
a small increase in the magnitude of the persistent currents for
very high disorder only \cite{schmitteckert98}. It has also been
shown that repulsive interactions may provoke significant disorder
and hence, enhance the electron mobility \cite{shepelyansky94,
benenti99} in few-particle models. The need to take into account
the spin has also theoretically been investigated using the method
of bosonization \cite{thierry95}. The result suggests that the
stiffness could be enhanced by intermediate interactions.
Numerical approaches \cite{roemer95,mori96} further show that
repulsive interactions enhance the persistent currents in weakly
disordered systems off half-filling only.

In this paper, we consider a simplified model of interacting
electrons in a one-dimensional disordered ring, which is
penetrated by a magnetic flux. It has been reported in another
study that moderate interactions could induce an enhancement of
the persistent currents even in the presence of weak disorder for
half-filled systems \cite{gambetti02}. Moreover disorder could
have the unexpected effect of increasing the zero-temperature
persistent currents in one-dimensional half-filled Hubbard rings,
with strongly interacting electrons. We emphasize that even off
half-filling the persistent current may still be enhanced by small
interactions and in such case, the filling may have an influence
on its behavior. Indeed, we show that the behavior of the
persistent currents changes as it is moved off half-filling, owing
to the appearance of holes. These results are established
partially within the framework of the strong coupling perturbation
expansion and partially through numerical calculations by means of
the Density Matrix Renormalization Group algorithm
\cite{DMRGbasics,white92}. Finally, we study the influence of the
spin on the behavior of the stiffness parameter and show that the
case of half-filling is rather particular. This manuscript
\cite{gambettithese04} is organized as follows. In Section
\ref{sec:modelandmethod}, the model and the calculation algorithms
of the persistent currents are presented. The numerical and
analytical work at half-filling are presented in section
\ref{sec:halffilling}. Section \ref{sec:awayhalffilling} contains
the numerical results off half-filling and further explains the
results by means of the perturbation expansion. Section
\ref{sec:spin} highlights the influence of the spin not only on
the stiffness parameter, but also on the magnitude of the
persistent currents. Section \ref{sec:localizationlength} is
devoted to the study of the localization length and identifies two
regimes that are characterized by weak and strong interactions,
which are separated by a crossover phase. Finally, our conclusions
are presented in section \ref{sec:conclusion}. Details of the
analytical approaches used in Sections \ref{sec:halffilling} and
\ref{sec:awayhalffilling} are given in Appendices
\ref{sec:analyticshf} and \ref{sec:analyticsaway}. A toy model is
developed for non half-filled systems in Appendix
\ref{sec:toymodel2on3}. Appendix \ref{sec:infUweightsGS} concerns
the behavior of the weights of the ground-state in the strong
interaction limit. In Appendix \ref{sec:oddnb}, the specific case
of an odd number of electrons present in a ring is discussed.

\section{\label{sec:modelandmethod} Model and Method}

A one-dimensional lattice is shown in function of the
Hubbard-Anderson Hamiltonian \cite{Hubbard57Anderson59}, which we
use to describe N interacting electrons in a disordered ring of M
sites:
\begin{eqnarray}
\label{eq:generalHamiltonian} H = H_{K} + H_{W} + H_{U}
\end{eqnarray}
\begin{eqnarray}
\label{eq:hk} H_{K}=-t \sum_{i=1}^{M}
\sum_{\sigma=\uparrow,\downarrow} \left(
c_{i,\sigma}^{\dagger}c_{i-1,\sigma}^{\phantom{\dagger}} +
c^{\dagger}_{i-1,\sigma}c_{i,\sigma}^{\phantom{\dagger}}\right)
\,,
\end{eqnarray}
denotes the kinetic energy, $t$ will be set equal to one in the
numerical part.
\begin{eqnarray}
\label{eq:hw} H_{W}=W
\sum_{i=1}^{M}\sum\limits_{\sigma}v_in_{i,\sigma}
\end{eqnarray}
stands for the disorder with $v_i\in[-\frac{1}{2},\frac{1}{2}]$,
$W>0$ and finally
\begin{eqnarray}
\label{eq:hu} H_{U}=U\sum_{i=1}^{M}n_{i,\uparrow}n_{i,\downarrow}
\end{eqnarray}
accounts for the electronic interactions.

The threading of the ring by a magnetic flux introduces an
Aharonov-Bohm phase $\Phi=2\pi\phi/\phi_0$ measured in units of
the flux quantum $\phi_0=h c/e$. According to Bayers and Yang
\cite{bayersyang} the only effect of the magnetic field is to
impose twisted boundary conditions
$c_{0,\sigma}=c_{M,\sigma}\exp{i\Phi}$ between the first and the
last site, the other pieces of the Hamiltonian remain unchanged.
The persistent currents that are carried through an isolated ring
are the direct consequence of the sensitivity of the eigenenergies
to a phase-change in the boundary conditions, in particular for
\begin{eqnarray}
\label{eq:currentdef} I(\phi) = -
\frac{\partial{E_0(\phi)}}{\partial{\phi}} \, ,
\end{eqnarray}
at vanishing temperature where $E_0(\phi)$ denotes the
ground-state energy. It has been shown that transport properties
are related to thermodynamic properties in a localized system at
vanishing temperature \cite{kohn64}. In fact, persistent current
are a periodic function of the flux $\phi$,
\begin{eqnarray}
\label{eq:Phidep} I(\Phi) = \left(E(0)-E(\pi)\right) \sin{\Phi}
\cdot \frac{\pi}{\phi_{0}} \,.
\end{eqnarray}
where
\begin{eqnarray}
\label{eq:phasesensitivity} E(0)-E(\pi)=\Delta E
\end{eqnarray}
is called phase sensitivity. We furthermore introduce the
stiffness, given by
\begin{eqnarray}
\label{eq:stiffness} D = \frac{M}{2}|\Delta E\,|,
\end{eqnarray}
which is related to the Drude weight $D_C$ \cite{kohn64}. We
compute $D$ because it is simpler to calculate
than the conductivity $\sigma$, and because it furthermore allows
for a measure of the magnitude of the persistent currents. The
energy levels are computed by means of the density matrix
renormalization group method DMRG \cite{white92,DMRGbasics} and
exact diagonalization. The first version of the program was
written by P. Brune \cite{brune} for the ionic Hubbard model.
Since $H$ is invariant under rotation around the $z$-axis its
eigenstates can be calculated by restricting it to the subspace
$S_{z}=0$. We used eight finite-lattice iterations and kept a
maximum of 700 states per block iteration.

\section{\label{sec:halffilling} Half-filling}

The case of half-filling is here considered when the number of
sites is equal to the number of electrons $M=N$.

The very strong interaction limit can be treated analytically in
order to better comprehend the behavior of the phase sensitivity
within the Mott insulator limit.  When $U \gg W,~t$ the
interaction $U$ dominates, and it is possible to expand $\Delta E$
in terms of $t/U$ \cite{bouchiath89,Dietmar01}. The details of the
calculation are given in Appendix \ref{sec:analyticshf}. The
basis-states $|\Psi_{\alpha} \rangle$ of $H_{0}$ spanning the
$\alpha$-space \cite{auerbach94} are given by the product of the
different on-site states, and read as follows:
\begin{eqnarray}
\label{basicstate}
|\psi_{\alpha}\rangle =
\left(\prod_{k=1}^{N/2}c_{i^{\uparrow}_k(\alpha),\uparrow}^{\dagger}\right)
\left(\prod_{k=1}^{N/2}c_{i^{\downarrow}_k(\alpha),\downarrow}^{\dagger}\right)
|0\rangle \, .
\end{eqnarray}
The function $i_k^{\uparrow (\downarrow)} (\alpha)$ specifies the
site where the $k^{\mathrm{th}}$-electron with spin $\uparrow$
($\downarrow$) among the $N/2 \uparrow(\downarrow)$-electrons is
created following the configuration $\alpha$. The $|0 \rangle$
state is composed of $M$ empty sites. The symbol
$|\Psi_{\beta}\rangle$ denotes states that do not possess
doubly-occupied sites (exactly one electron per site is present at
half-filling), spanning the $\cal{S}$-space;
$|\Psi_{\gamma}\rangle$ denotes states that contain at least one
doubly-occupied site spanning the $\cal D$-space. The
corresponding many-body energies of the states
$|\psi_{\alpha}\rangle$ are summed over the disorder contributions
$E_{\alpha}^{\rm W}$ and the interaction energies $E_{\alpha}^{\rm
U}$. They are equal to:
\begin{eqnarray}
    \label{eq:basisenergies}
    E_{\alpha} = & W \sum_{k}
\left(v_{i^{\uparrow}_k(\alpha)}+v_{i^{\downarrow}_k(\alpha)}\right) + & U g_{\alpha} \, .
\end{eqnarray}
Here $g_{\alpha}$ stands for the number of doubly-occupied sites.
In the Mott insulator limit when $U \gg W,t$ the $|\Psi_{\gamma}
\rangle$'s possess a very high energy level and hence, a gap
arises with the $|\Psi_{\beta} \rangle$'s.

First, we consider the case without disorder. The $|\Psi_{\beta}
\rangle$'s are the basis-states and all possess an equal zero
energy level. Hence, for very large $U$, $H=H_0+H_{W}$ and the
kinetic energy term $H_{K}$ represents a perturbation with respect
to the dominating term $H_{U}$. Within the limit of $U \gg t$, the
spin degree of freedom can be treated as a significant spin
Hamiltonian that is in fact the antiferromagnetic Heisenberg
Hamiltonian when $\phi=0$, for the case of the clean system
\cite{lieb62,lieb93,auerbach94}. The antiferromagnetic Heisenberg
Hamiltonian ground-state is the superposition of all states with a
total spin equal to zero. It is furthermore non-degenerate. This
theorem has been extended to the half-filled Hubbard model
\cite{lieb62,lieb93} for an even number of sites $M$
\cite{lieb93,tasaki95} and the ground-state can be expressed as:
\begin{eqnarray}
\label{eq:exactGShalf}
|\psi_0\rangle =\sum_\beta f_\beta
\left(\prod_{k=1}^{N/2}c_{i^{\uparrow}_k(\beta),\uparrow}^{\dagger}\right)
\left(\prod_{k=1}^{N/2}c_{i^{\downarrow}_k(\beta),\downarrow}^{\dagger}\right) |0\rangle
\end{eqnarray}
where the $0 < f_{\beta} < 1$, according to Marshall's theorem
\cite{marshall55}, for all spin configurations $\beta$ in the
subspace $\cal S$ and $i_k^{\uparrow} \ne
i_{k^{\prime}}^{\downarrow} \forall k \ne k^{\prime}$. The
ground-state for $\Phi = \pi$ is equal to that presented in eq.
(\ref{eq:exactGShalf}).

Now we calculate the leading contributions to the phase
sensitivity $\Delta E$ using the difference present in the
higher-order corrections of the many-body ground-state energies
under periodic and antiperiodic boundary conditions. Studying the
Hubbard Hamiltonian within the strong interaction limit, is
equivalent to studying the action of $H_K$ in the $\cal S$-space
\cite{tasaki95}. Indeed, the numerator of $E^{(n)}$ is composed of
matrix elements
$\langle\psi_{\gamma_l}|H_{K}|\psi_{\gamma_{l+1}}\rangle$ of the
hopping part of the Hamiltonian, which connects each site to its
neighbor. Thus, the resulting correction brought to the
ground-state energy $E^{(n)}$ is different from zero only when all
$|\psi_\gamma\rangle$ states can be connected by one-particle
hopping processes \cite{Dietmar01,selva00}. These {\it sequences}
are called  ${\bf A}^{(\beta,\beta')}$ and start at
$|\psi_{\beta}\rangle$ and end at $|\psi_{\beta'}\rangle$, with
the intermediate states being all $|\psi_{\gamma}\rangle$'s.
In order to calculate the difference between periodic and
antiperiodic boundary conditions, the particles must circle the
ring and cross the border once. The lowest contributions are of
order $M$, and in such case, the energy corrections for periodic
and antiperiodic conditions differ only  by the sign. Intermediate
results are given in an already published article
\cite{gambetti02}.

Second, we reintroduce disorder. The $\cal S$-states follow
Mott-insulator configurations and all possess a similar and lowest
energy level that reads $E_0^{W}=W\sum_{i=1}^{M=N}v_i$ (from eq.
(\ref{eq:basisenergies})). This energy level is independent of the
spin configurations of the $\cal S$-states and consequently,
within this limit, they degenerate and are characterized by the
same weights $f_{\beta}(U \to \infty)$. All the terms in the
denominator are calculated by differences between disorder and
interaction contributions of the ground-state $|\Psi_{0}\rangle$
and the intermediate state $|\Psi_{\gamma_l}\rangle$. The
denominators are written as $Wd_{\gamma_l}+g_{\gamma_l}U$. We
further define $(E_{\gamma_l}^{\rm W}-E_0^{\rm W})=Wd_{\gamma_l}$,
as the energy difference between the disorder configurations of
the state $|\Psi_{0} \rangle$ and the state
$|\Psi_{\gamma_l}\rangle$. We develop $\Delta E$  in powers of
$\frac{W}{U}$ up to second order. This yields :
\begin{eqnarray}
\label{eq:Wexpansionfinal}
\Delta E^{(M)}\approx\frac{
  (-1)^{N/2}4t^{M}}{U^{M-1}}\sum_{\beta,\beta'}\sum\limits_{{\bf A}_{\rm f}^{(\beta,\beta')}}
\frac{f_\beta f_{\beta'}}{\prod_{l}g_{\gamma_l}} \nonumber\\
\left(1+\frac{W^2}{U^2}
\left(\sum_l\frac{d_{\gamma_l}^2}{g_{\gamma_l}^2}
+\sum_{l<m}\frac{d_{\gamma_l}d_{\gamma_m}}
                {g_{\gamma_l}g_{\gamma_m}}\right)\right).
\end{eqnarray}
This sum is performed over all forward hopping sequences ${\bf
A}_{\rm f}^{(\beta,\beta')}$ (for the electrons that are moving in
the anti-trigonometric direction of motion) with a first factor 2
introduced for the backward sequences and an additional factor 2
accounting for the difference between periodic and antiperiodic
contributions. All contributing sequences lead to a cyclic
perturbation of the $N/2$ operators corresponding to electrons
with a specific spin direction. Since the weights $f_\beta$ are
all positive, the sign of the phase sensitivity at strong
interaction is given by $(-1)^{N/2}$ in eq.
(\ref{eq:Wexpansionfinal}), as in the non-interacting case
\cite{Leggett}. The dominating term in eq.
(\ref{eq:Wexpansionfinal}) (corresponding to the clean limit $W
\rightarrow 0$) exhibits an interaction-induced suppression of the
persistent current $\propto U(t/U)^M$. The first-order correction
in $W/U$ totally vanishes due to the particle-hole symmetry. The
second-order correction is positive because the first term is
positive, and much larger than the second one. We conclude that
the phase sensitivity has positive corrections for order
$(\frac{W}{U})^2$. This conclusion is also true for the stiffness
parameter, as $D=\frac{M}{2}|\Delta E|$. The unexpected conclusion
is that at half-filling, {\it the disorder increases both the
phase sensitivity and the magnitude of the persistent currents for
highly interacting electrons in Hubbard-Anderson rings}
\cite{gambetti02}. This counterintuitive effect
\cite{lee85kramer93} might also be related to a competition
between the phenomenon, disorder and interaction. The ground-state
energy, which is equal to $E_0 = W \sum_i v_i$ increases with the
strength of the disorder W, and when $W \le U$, the energy level
becomes comparable to the interaction strength $U$. Under these
conditions, the gap reduces between the energy levels of the
excited states $|\Psi_{\gamma}\rangle$ and the ground-state. A
jump to a doubly-occupied site is then rendered easier and thus,
the movement of the electrons around the ring is favored,
enhancing the magnitude of the persistent currents. These sets of
$M$ hops illustrate a simple mechanism that can explain the
changes in the magnitude of the persistent currents under strong
interaction limits.

Extensive numerical calculations were performed for systems of
$N=14$ electrons ($7\uparrow; 7\downarrow$) on $M=14$ sites, and
for a sample of $N=20$ particles ($10\uparrow; 10\downarrow$) on
$M=20$ sites.

A selection of the probability distributions of $D(U)$ are plotted
for fixed $M$ and $N$ values, for set disorder and interaction
values. They can be fitted by Gaussian functions and are said to
be log-normal.
\begin{figure}[h!]
\includegraphics*[width=0.4\textwidth]{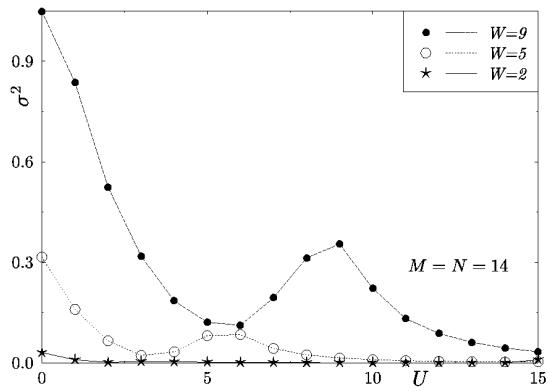}
\caption{Variance of the logarithm of the stiffness as a function
of $U$, for $W=2,5,9$ and for $M=N=14$.}
\label{fig:sigma1414}
\end{figure}
In Fig. \ref{fig:sigma1414}, the variance $\sigma^{2}$ is plotted
as a function of $U$, for $M=N=14$. For $U=0$, the fluctuations
are governed by the disorder and they increase when disorder
increases. When interactions are introduced, fluctuations of two
different natures compete. Figure \ref{fig:sigma1414} illustrates
this by showing that these fluctuations compensate each other and
decrease, phenomenon that leads to an increase in the magnitude of
the persistent currents. Overall, this produces a maximum of
magnitude for the persistent currents around $U \sim W/2 \pm t$.
For moderate $U$, the particles can move and two electrons can be
together on the same site. Thus, the electrons explore sites with
higher on-site potentials and under such conditions, the
fluctuations are more sensitive to disorder. Finally, in the
region where $U \lesssim W$, the variance increases. After this
second maximum, the fluctuations decrease when the interaction
increases for $U>W$. Within the region of very strong
interactions, the particles are trapped alone within their sites
and thus, their movement is frozen with exactly one particle per
site. Fluctuations are on-site only and decrease systematically.

\begin{figure}[h!]
    \includegraphics*[width=0.4\textwidth]{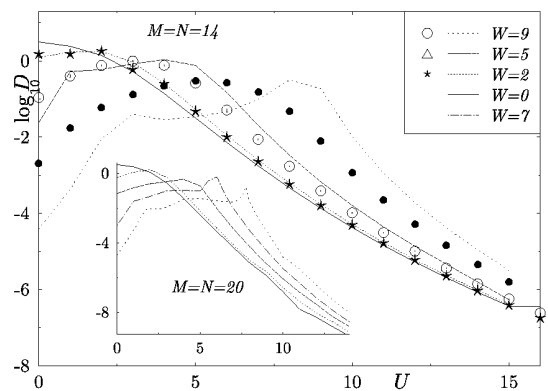}
\caption{Dependency of the stiffness parameter upon the degree of
interaction, for $M=N=14$ ($M=N=20$ in the inset), for
disorders $W=0,2,5,9$.  Lines represent individual samples.
Symbols correspond to the global average. Error bars are smaller
than the symbols size and hence, do not appear. Note that the
curve for $W=0$ behaves as a Luttinger liquid. }
    \label{fig:1414}
\end{figure}
Fig. \ref{fig:1414} shows the dependency of the stiffness
parameter $\log{D}$ on the degree of interaction, for $M=N=14,20$
for typical individual samples. These samples are represented by
the single lines and global means are illustrated for three
different disorder values, $W$, and for the clean case where $W=0$
(solid line). The ensemble averages of $\log{D}$ are performed
over approximately 100 different samples (disorder realizations).
The behavior of both the samples and the means are similar but in
contradiction with the spinless fermions \cite{Dietmar01}. Indeed,
our curves are all rather smooth \cite{ramin95,thierry95} even if
small peaks appear for the individual samples for strong disorders
($W \sim 7,9$) in Fig. \ref{fig:1414}. This second peak increase
of $\log{D}$ for samples occurs for $U \sim W-t$ when the
fluctuations are strong (cf. Fig. \ref{fig:sigma1414}). For $U=0$,
the disorder leads to the Anderson localization and $\log{D}$ is
strongly suppressed by increasing disorder \cite{lee85kramer93}.
This is what would be expected in one dimension, for a
non-interacting disordered system. For clean rings ($W=0$), the
interaction always reduces the stiffness, which is consistent with
a Luttinger liquid calculation \cite{loss92,stafford90}.

\begin{figure}[h!]
    \includegraphics*[width=0.4\textwidth]{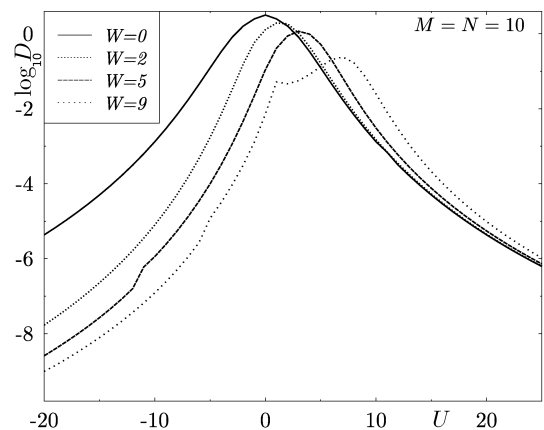}
\caption{Dependency of the stiffness parameter for an individual
sample upon the degree of interaction $U$, for $M=N=10$ and for
disorders $W=0,2,5,9$. Interactions are both attractive and
repulsive, $-20<U<25$. An increase occurs for positive
interactions only. }
    \label{fig:negU1010}
\end{figure}

A weak repulsive interaction $U>0$ leads to an increase of the
stiffness compared to its value for $U=0$, even when little
disorder is present; such an increase was in fact predicted from a
renormalization group approach, for both off half-filling and
moderate disorders \cite{thierry95}. Indeed, the stiffness is less
sensitive to disorder when repulsive interactions are accounted
for than when attractive interactions only are considered, off
half filling and for moderate disorders. This is clearly
illustrated in Fig. \ref{fig:negU1010}, for an individual sample
$M =N=10$. For attractive interactions no increase occurs and the
stiffness phenomenon is strongly suppressed. The curves
corresponding to different disorders do not cross to negative
interactions, in accordance with the analytical development of
Sec. \ref{sec:halffilling}, which is valid only for positive
interactions \cite{lieb93}.
At intermediate interaction levels i.e. for $U \in [t, W]$, the
phase sensitivity exhibits a maximum around $W/2 \pm t$, which
becomes broader with increasing disorder. When disorder increases
this maximum decreases, but the ratio $D_{\mathrm max}/D(0)$ is
greater. With $M=N=14$, the mean ratios are equal to $1.4$,  $8.1$
and $150$ for $W=2,5$ and $9$, respectively. These ratios are even
greater for samples with $M=N=20$ and can under certain conditions
be three times greater (e.g. for $W=9$).

\begin{figure}[h!]
\includegraphics*[width=0.4\textwidth]{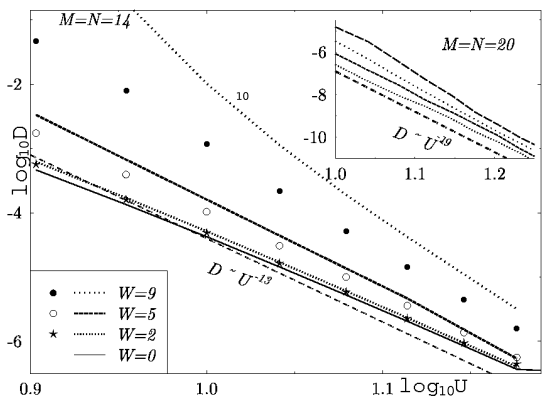}
\caption{Data are plotted on log-log scale for $M=N=14$ ($M=N=20$
in the inset) under strong interaction conditions. Results
confirm the power law $D \propto U^{-(M-1)}$. Symbols illustrate
global means and lines individual samples.} \label{fig:log1414}
\end{figure}

For large repulsive interaction $U\gg W$, the behavior of
$\log{D}$ is radically different. The stiffness decreases strongly
with the degree of interaction and the numerical results confirm
this with the analytical power law $\propto 1/U^{M-1}$, as
illustrated in Fig. \ref{fig:log1414}. The slopes for moderate
interactions are stronger when the disorder strength W increases
in order to recover the $1/U^{M-1}$ behavior when $U \gg W$.
Results for clean systems ($W=0$) and strong interactions gave the
means to check a good concordance between observed results and the
power law. This decrease is related to the localization due to the
interaction \cite{loss92}. The most surprising result is that a
curve crossing occurs within the Mott insulator limit (see Fig.
\ref{fig:1414}) confirming that the disorder induces an increase
in the stiffness parameter, as found analytically.

For all values of $U$ and $W$, we obtained negative $\Delta E$ for
$M=N=6,10,14$, and positive $\Delta E$ for $M=N=20$. Thus,
consistent with both Leggett's rule \cite{Leggett} for the
non-interacting case and the analytical result for very strong
interactions, the sign of $\Delta E$ is determined by the number
of electrons that are each characterized by a given and specific
spin, according to $(-1)^{N/2}$. Our numerical results revealed
that this rule is valid for all tested samples and for all
considered interaction and disorder values. In the following
section, these results are compared to the phenomenon observed off
half-filling.

\section{\label{sec:awayhalffilling} Off Half-Filling}

This section concerns systems off half-filling, $N\ne M$. The very
strong interaction limit can be treated analytically and we
maintain the notations presented in Sec. \ref{sec:halffilling}.
Details of the calculations are given in Appendix
\ref{sec:analyticsaway}. An example is developed in a toy model in
Appendix \ref{sec:toymodel2on3}. The expansion is valid for an
{\it even} number of particles and a {\it single} empty site,
called a hole \cite{lieb93,tasaki95}.

Once again, we first consider the case without disorder. For very
large $U$ the kinetic energy term $H_{K}$ represents a
perturbation with respect to the dominating term $H_U$ as at
half-filling \cite{tasaki95}. We must first determine the
ground-state of the system. The basis-states are still given by
eq. (\ref{basicstate}). Now since a hole has been introduced, the
$\cal S$-states are connected by only one hop. A given basis-state
$|\Psi_{\beta}\rangle$ can be connected to itself within an
ensemble of first order hops, which are all going in the same
direction to form a super-lattice \cite{nagaoka66}. For the toy
model of two electrons on three sites of Appendix
\ref{sec:toymodel2on3}, the number of super-lattices is equal to
one. Increasing both the number of sites and the number of
electrons increases the number of super-lattices and leads to the
creation of $N/2$ subblocks, which is the number of electrons with
the same spin present within the ring. We call these subblocks
${\cal C}_i$, where $i$ is the number of electrons of similar spin
ones besides the others.
Each subblock ${\cal C}_i$ leads to a different eigenvector, which
implies that the ground-state degenerates. The basic structure of
the ground-state can be determined by the standard
Perron-Frobenius sign convention \cite{perronfrobenius} in the
case of a single empty site $N=M-1$. The ground-state is in such
case a superposition of the $|\Psi_{\beta}^{{\cal C}_i} \rangle$
within each subblock ${\cal C}_i$. The sign of the weights depends
on the boundary conditions \cite{tasaki95,lieb93}. If the
transition amplitude between two states is negative, then one can
superimpose within a subblock, states with the similar signs.
Viceversa, if the transition is positive, one can superimpose
states with opposite signs. These are in fact the Marshall sign
rules \cite{marshall55} that are extended to the Hubbard model
\cite{weng91}. The ground-state inside each subblock is written
as:
\begin{eqnarray}
    \label{eq:GSAHFW=0}
    |\Psi_{0}^{{\cal C}_i}\rangle=\sum_{\beta}f_{\beta}^{{\cal C}_i}|\Psi_{\beta}^{{\cal
C}_i}\rangle \left \{
     \begin{tabular}{c}
        $f_{\beta}^{{\cal C}_i} > 0$ for $\Phi = 0$ \\
        $f_{\beta}^{{\cal C}_i} \in \Re$ for $\Phi = \pi$
    \end{tabular}
    \right .
\end{eqnarray}
where the weights $f_{\beta}$ are chosen to be real
\cite{marshall55,perronfrobenius}. Thus, for antiperiodic boundary
conditions within each subblock ${\cal C}_i$, half of the
basis-states will possess a positive (or negative) sign
\cite{lieb93,tasaki95}. The first order corrections of the
many-body ground-state energies are equal for periodic and
antiperiodic boundary conditions. Indeed, in case of antiperiodic
boundary conditions, each time an electron crosses the border, the
sign due to the flux is compensated for by the sign of the
weights. The phase sensitivity cancels out at first order. At
second order, the intermediate states are obviously equal to
$|\Psi_{\gamma} \rangle$'s, with exactly one doubly-occupied site.
Lets now consider those sequences that cross the border. For
example, that sequence where a particle jumps to form a
doubly-occupied site and then, one of these electrons comes back
to the initial site or jumps onto a neighboring site. These
sequences are called exchange and double-jump. Their relative
terms, when considering the two Hamiltonians, have opposite signs
for periodic and antiperiodic boundary conditions. The eigenvalues
for these matrices are different and hence, the phase sensitivity
reads:
\begin{eqnarray}
    \label{eq:deltae2away}
    \Delta E^{(2)}=& |E^{(2)}(0)-E^{(2)}(\pi)|\nonumber\\
    \Delta E^{(2)}=& \frac{-t^{2}}{U} \kappa
\end{eqnarray}
where the factor $\frac{-t^{2}}{U}$ appears in all parts of both
matrices. The phase sensitivity behaves as $1/U$ within the very
strong interactions limit.

Now, if one introduces disorder, then the particles stay on the
sites with the lowest disorder potentials inside each subblock
${\cal C}_i$ and form the ${\cal C}_{i}^{R}$-subspace. But one
maintains $N/2$ subblocks. Hence, the ground-state is now a
superposition of the basis-states ${\cal C}^{R}_{i}$ with the
weights $f_{\beta}$ respecting the Marshall rule
\cite{marshall55,perronfrobenius}. A single jump does not lead to
a basis-state anymore and one considers it now as two jumps. Under
such conditions, the only possible sequences are the exchanges
that lead to opposite values for periodic and antiperiodic
boundary conditions. The phase sensitivity at second order here
reads, after an expansion of $W/U$:
\begin{eqnarray}
    \label{devorder2}
    \Delta E^{(2)}= \frac{-2t^{2}}{U} \kappa^{\prime}
(1+\frac{W^{2}}{U^{2}}d_{\alpha}^{2})
\end{eqnarray}
This $1/U$ power law is lower than at half-filling. Indeed, for
$N=M$, the ground-state is rigid. When $N < M$, with the provided
hole, the electrons can move more easily around the ring, which
decreases the power law compared to the half-filled situation.
This nicely illustrates the behavior of the persistent currents
off half-filling under strong interaction limits. Electrons move
around the ring without meeting each other and when close to the
border, they hop on a doubly-occupied site. The phase sensitivity
then decreases as the disorder strength $W$ increases.
Unfortunately, the perturbation theory does not allow us here to
conclude about the sign of $\Delta E$ under strong interaction
limits.

We now present numerical simulations for $N=8$ particles
($4\uparrow; 4\downarrow$) on $M=9$ and on $M=11$ sites. The
symbols stand for the mean performance calculated over 100
disorder realizations (within a given sample); the lines represent
individual samples. The results concentrate on an even number of
electrons $N$, and we specifically consider the effect of the
parity of the number of sites $M$. For an odd and an even value of
$M$, the behavior of the stiffness is similar. Moreover, samples
present similar behaviors as that observed for the mean curves.
Finally, all curves are relatively smooth. The averages are
calculated over the logarithm of the stiffness, $\log{D}$.

\begin{figure}[h!]
\includegraphics*[width=0.4\textwidth]{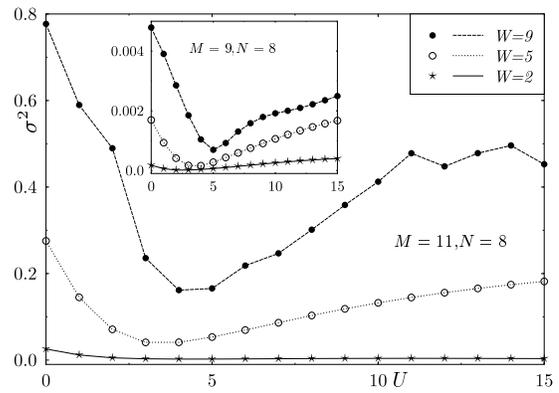}
\caption{Variance of the logarithm of the stiffness $\log{D}$ as a
function of $U$ for different strengths of the disorder $W=2,5,9$
for $M=11$ ($M=9$ in the inset) and $N=8$.}
\label{fig:sigmaaway}
\end{figure}

The variances for three values of the disorder $W=2,5,9$ are
plotted as a function of the interaction for $N=8$ particles, for
$M=9$ and $M=11$ sites. Results show that they are {\it similar}
in Fig. \ref{fig:sigmaaway}.
For $0<U<W/2 \pm t$, the behavior is similar to that expose in
Sec. \ref{sec:halffilling}, with the fluctuations that decrease.
The persistent currents reach a maximum around $W/2 \pm t$.
Considering the strong interaction limits, the particles are
trapped alone within their sites but a movement is still permitted
because of the existence of an empty site (hole). The system is
thus not rigid. Fluctuations are more significant than at
half-filling since the fluctuations of both nature do not
compensate, i.e. the on-site fluctuations and the fluctuations
that are due to the disorder present on the empty site. When the
disorder increases the fluctuations in the empty site increase,
and so does $\sigma^{2}$.

\begin{figure}[h!]
\includegraphics[width=0.4\textwidth]{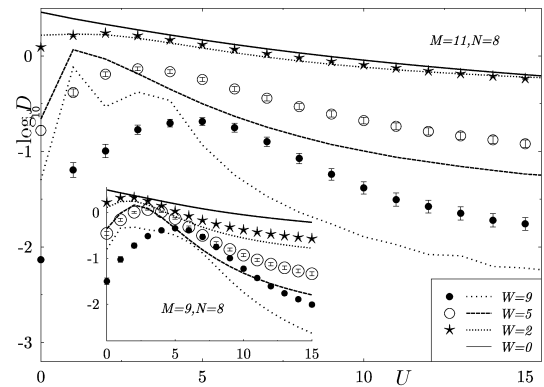}
\caption{Dependency of the stiffness upon the degree of
interaction $U$, for $N=8$, $M=11$ ($M=9$ in the inset) and
$W=0,2,5,9$. Lines represent $\log{D}$ for individual samples
and symbols correspond to global means.}
    \label{fig:118}
\end{figure}

In Fig. \ref{fig:118}, the logarithm of the stiffness $\log{D}$ is
plotted as a function of the interaction $U$, for different values
of the disorder $W$ and for the clean case specifically.
At $U=0$ $\log{D}$ is strongly suppressed by the disorder
(Anderson localization) \cite{lee85kramer93}. For clean rings, we
verify that the interaction systematically reduces the stiffness,
which is consistent with a Luttinger liquid calculation
\cite{loss92}. A weak repulsive interaction leads to an increase
of the stiffness respectively to its value in absence of
interactions, as predicted by a renormalization group for moderate
disorders \cite{thierry95}. We obtain this increase for all
fillings and all non-zero disorder strengths. This increase can
reach factors equal to approximately 1.5 (1.3), 4.5 (3.4), 28 (17)
and when $W=2,5$ and $9$, respectively. This is true for the
average system $M=11$ ($M=9$) and $N=8$. The increases of the
samples are important and can reach an order of magnitude.

\begin{figure}[h!]
\includegraphics*[width=0.4\textwidth,angle=0]{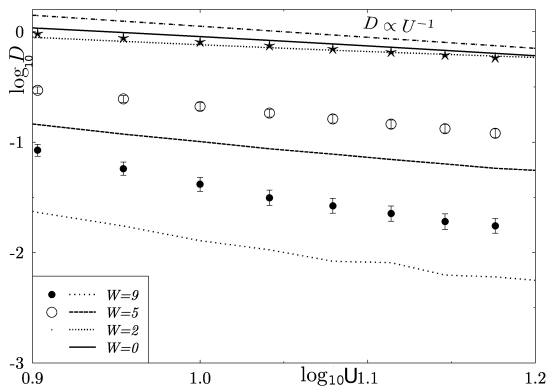}
\caption{Data are plotted for strong interaction conditions in
log-log scale for $M=11$ and $N=8$. They reveal a power law of
$1/U$. The symbols represent means and the lines, individual
samples.}
\label{fig:log118}
\end{figure}

\begin{figure}[h!]
    \begin{center}
\includegraphics*[width=0.4\textwidth]{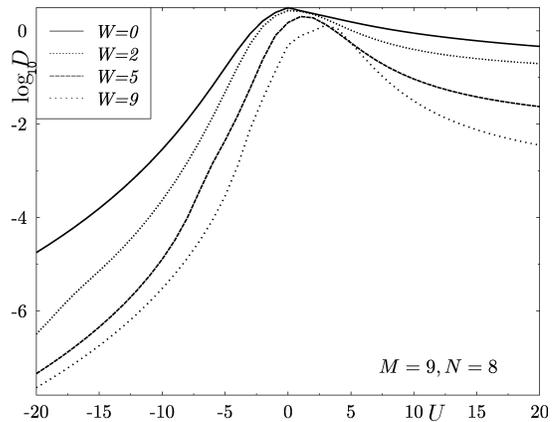}
    \end{center}
    \caption{Illustration of the behavior of $\log D$ for a sample of $M=9$ sites and $N=8$
electrons for negative and positive interactions $-20<U<20$.
Note that an increase occurs for positive interactions only.}
    \label{fig:negU98}
\end{figure}

The stiffness $\log{D}$ decreases as the interaction $U$ increases,
within the very strong interaction limits.
Figure \ref{fig:log118} presents the logarithm of the stiffness
$\log D$ as a function of $\log U$ for $M=11$, $N=8$ and gives
the means to determine that the power law follows $1/U$. We have
verified that for the system $M=9$,$N=8$, the stiffness decreases as a function of $U^{-1}$,
in agreement with strong interaction situations. For important
disorder conditions, the slope must be calculated for interactions $U \gg W$
in order to recover the $1/U$ law.
This law is true for several holes even if the perturbation theory
is valid only for single holes. In Fig. \ref{fig:negU98}, the
stiffness is smaller for negative interaction values than for
positive ones \cite{thierry95}, and no stiffness increase occurs.

The sign of the phase sensitivity was also studied and results
demonstrate that it also follows the Leggett's rules
\cite{Leggett}. The sign depends on the number of electrons that
possess a specific spin, according to $(-1)^{N/2}$. This was
observed for all disorder and interaction values. The present
result was verified numerically for all the computed samples. For
$N=8$, the phase sensitivity is always positive and for $N=6$ the
phase sensitivity is negative, whatever the number $M$ of sites.

\section{\label{sec:spin} Role of the spin}

The logarithm of $D$ gave us the means to model the behavior of
the persistent currents as a function of $U$, for different values
of the disorder $W$ and for different system sizes $M$. Our
conclusion is that a pronounced increase occurs for all degrees of
disorder not equal to zero, for every system size $M$ and
arbitrary fillings, but for moderate values of the interaction
$U\sim W/2$.

This increase can be explained through the competition between
disorder and interaction phenomenon. Indeed, the disorder sets the
electrons within the sites with the lowest potentials whereas the
interaction sets a unique electron per site. When disorder and
interaction are of the same strength, the determination of the
ground-state of the system is difficult. Indeed, the electrons
oscillate between being together on the same site hence, adding an
energy level to $U$ or being on two different sites which sets a
different degree of disorder, energy $Wv_i$. However, the
ground-states directly depend on the spin parameter, where the
case of half-filling corresponds to $M=N$. When $W>U$, electrons
may prefer avoiding the site with highest potentials and hence,
will generate a doubly-occupied site (a spin up and a spin down).
When $U$ and $W$ are of the same order, the electrons move easily
as the doubly-occupied site can hop freely around the ring and set
itself next to the site with the highest potential. Under such
conditions, if $W\sim U$, a particle from the doubly-occupied site
can jump on the site and further on to its neighbor, recreating a
doubly-occupied site. The quantity here important to consider is
the difference, $U-W/2$. When it is smaller than the amplitude
transition $t$, the electrons movement around the ring is easy and
induces an important current. This behavior can be related to
calculations for 2D spinless fermions with disorder and long range
Coulomb interaction \cite{benenti99}. When $W<U$, the particles
remain alone on their respective sites and this leads to a
decrease in the magnitude of the current. For off half-filling,
this explanation is also valid but the holes are placed upon the
highest disorder potentials. Hence, the maximum current further
decreases. This simple mechanism of two competing ground-states is
a starting point for a better understanding of what happens in two
dimensions, and provides a starting indication of how persistent
currents may be enhanced by large orders of magnitudes.

\begin{figure}[h!]
   \begin{center}
       \includegraphics*[width=0.4\textwidth]{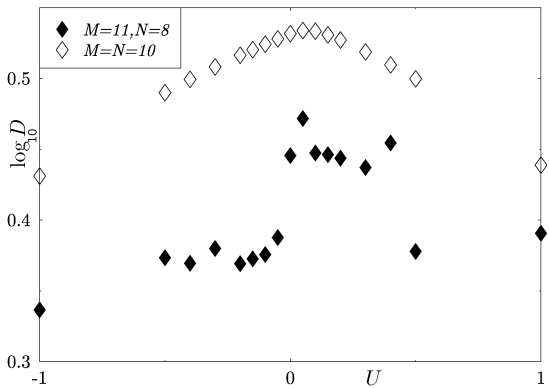}
   \end{center}
   \caption{Interaction dependency of the stiffness parameter on the degree of interaction, for $W=0.1$ in
   the case of two systems, $M=N=10$ and $N=8$,$M=11$. These results are calculated from a group average
   of 100 samples.}
   \label{fig:w=0.1}
\end{figure}

We illustrate this modeling considering first a very weak disorder
$W=0.1$, for the two following systems $M=N=10$, and $N=8$,
$M=11$. Here $\log{D}$ is calculated from a group average of about
100 samples. An increase is already observable for such a weak
disorder in Fig. \ref{fig:w=0.1}. One can see that a maximum
increase is obtained for $U \sim 0.05=W/2$.
These results give us the means to conclude to an increase of the
stiffness for non-vanishing disorders. The persistent currents are
less important for $M=11$, $N=8$ than for $M=N=10$.

For negative interactions, the stiffness always decreases. Indeed,
two electrons with opposite spin directions can both generate a
doubly-occupied site and be on the sites with lowest potentials.
This decreases the stiffness as soon as $U<0$ for all disorders
strengths, as found in Fig. \ref{fig:w=0.1}.

\begin{figure}[h!]
    \begin{center}
        \includegraphics*[width=0.4\textwidth]{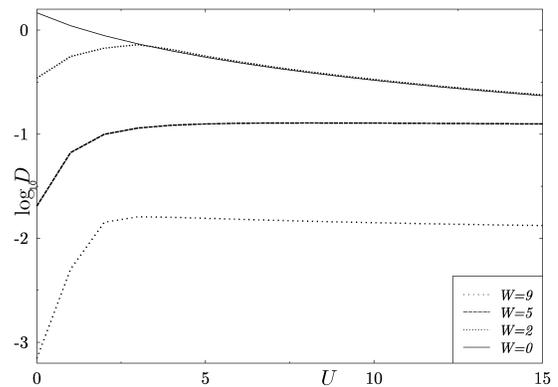}
    \end{center}
    \caption{$\log{D}$ is plotted as a function of $U$, for different values of the degree of disorder $W$,
    for an individual sample of $N=4$ electrons present on $M=13$ sites.}
    \label{fig:134}
\end{figure}

In Fig. \ref{fig:134}, a sample of $N=4$ electrons on $M=13$ sites
is presented in order to show that even for low fillings i.e.
$(N/M)\sim 0.3$, persistent currents increase. This increase
reaches an order of magnitude for moderate ($W=5$) and strong
($W=9$) degrees of disorder.
Thus, the mechanism that have been presented above to explain the
increase of the persistent current is valid even for low fillings
and for attractive and repulsive interactions in presence of
disorder. Other studies have previously considered the role of the
spin and have also concluded that it is an important parameter for
both one and two dimension situations, and for multi-channelled
systems \cite{georges94b}. The present work shows in addition that
in one dimension, the spin has a significant influence on the
magnitude of the persistent currents.

In the previous sections (\ref{sec:halffilling} and
\ref{sec:awayhalffilling}), the limit for the strong interactions
revealed different behaviors for half-filled and non half-filled
systems.
\begin{figure}[h!]
    \begin{center}
        \includegraphics[width=0.4\textwidth]{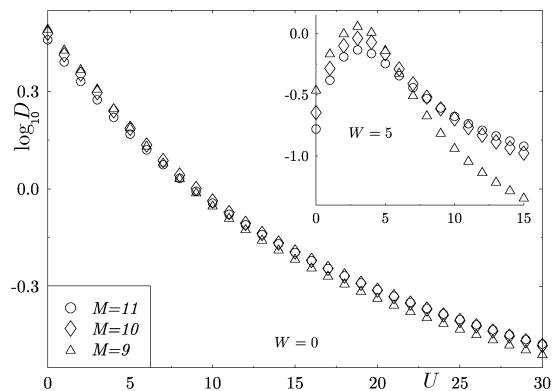}
    \end{center}
    \caption{$\log{D}$ is illustrated as a function of $U$, for $W=0$ ($W=5$ in the inset)
    for systems with $N=8$ and $M=9,10$ or $11$. A significant crossing between the curves is
observed for strong interactions. }
    \label{fig:dwfixaway}
\end{figure}
We now consider the case of a similar number of particles and a
similar degree of disorder (even $W=0$) but with different filling
properties (i.e. not equal to 0.5). For strong interactions i.e.
when the filling decreases for a given number of electrons, the
stiffness increases. This is illustrated in Fig.
\ref{fig:dwfixaway}. The presence of holes makes the persistent
currents increase within the strong interaction limit. The case of
half-filling deserves to be treated apart and presents a different
behavior, which underlines the important effect of the
particle-hole symmetry for the electrons \cite{denteneer01}.

\section{\label{sec:localizationlength} Localization Length}

In this final section, we study the localization length $\xi$ in
order to characterize the finite-size effect \cite{lee85kramer93}.
In order to extract $\xi$, we use a relation that is valid for
localized systems. The localization length is related to the
stiffness $D$ through:
\begin{eqnarray}
    \label{eq:loclexpression}
    \ln{D}=\ln{(A(U))}-M/\xi(U).
\end{eqnarray}
$\log{D}$ is plotted as a function of $M$ and one must check that a
straight line is obtained. Thanks to a linear regression the
slope, which is proportional to $1/\xi(U)$ can be obtained from
the numerical data of $\log{D}$, indicated here by the dimension of
the error bars. This relation is phenomenological and is obtained
thanks to dimensional reasoning but does not allow for the
determination of the sign.

\subsection{\label{ssec:loclengthhf} Localization length at half-filling}

\begin{figure}[h!]
    \begin{center}
        \includegraphics[width=0.4\textwidth]{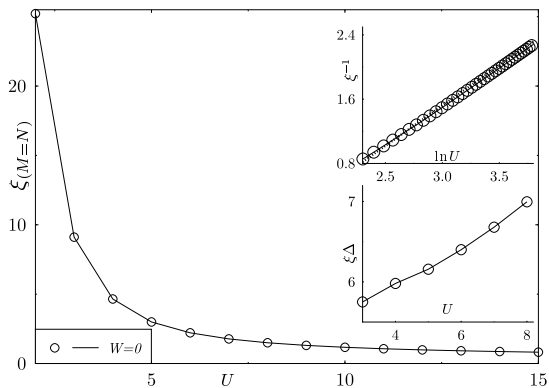}
    \end{center}
    \caption{$\xi$ is plotted as a function of $U$, for $W=0$.
    The symbols represent the values of $\xi$ obtained from $\log{D}$.
    In the bottom inset, $\xi \Delta$ \cite{LiebWu} is presented as a function of the interaction $U$, for $3<U<8$.
    In the top inset, $\xi^{-1}$ behaves according to the Bethe Ansatz within the strong interaction limits
    \cite{stafford90}.}
    \label{fig:xiW0}
\end{figure}

Disordered non-interacting one-dimensional systems are Anderson
insulators \cite{lee85kramer93}, whereas a half-filled system
behaves as a Mott insulator for very strong interactions
\cite{Mott}. We consider half-filled rings $M=N=6,10,14$.
We first verified the results obtained for $W=0$. First, plots
representing $\log{D(M)}$ must be straight. For $U=W=0$, a free
electron gas is obtained and the localization length should both
diverge and be infinite. Because of the numerical errors, one gets
a very high finite value of the localization length $\xi$. Once
the interaction is introduced, $\xi$ can be extracted. It
decreases as a function of the interaction increase (see Fig.
\ref{fig:xiW0}). This has previously been reported
\cite{stafford90}.
For moderate interactions, $U\in[3,8]$, the quantity $\xi\Delta$
($\Delta=\frac{8t^{2}}{U} \int_{1}^{\infty}
\frac{\sqrt{(y^{2}-1)}dy}{\sinh(2\pi ty/U)}$, the Lieb-Wu charge
gap \cite{LiebWu}) must reach the mean-field result of $(\xi
\Delta)/t=4$ \cite{stafford90}. The values calculated thanks to
our numerical approach reached 4.9, for $U=0$ (see the bottom
inset of Fig. \ref{fig:xiW0}). For very strong interactions the
inverse of the localization length behaves as the analytical
solution given by the Bethe Ansatz \cite{stafford90}. This is in
agreement with what was presented in Sec. \ref{sec:halffilling}.
The inverse of the localization length $\xi^{-1}$ (see the top
inset of Fig. \ref{fig:xiW0}) behaves as $-1.49+\ln{U}$ for $W=0$.

\begin{figure}[h!]
    \begin{center}
        \includegraphics[width=0.4\textwidth]{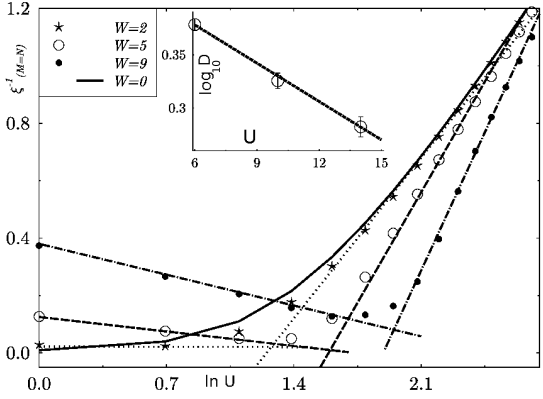}
    \end{center}
    \caption{$\xi^{-1}(M=N)$ is plotted in function of $\ln{U}$.
    The dotted lines represent the fitting of both regimes for $W=2,5$ and $9$.}
    \label{fig:unsurxilnuhf}
\end{figure}

We want to study the influence of the competition between disorder
and interaction on the localization length $\xi$. The inverse of
$\xi$ is represented in Fig. \ref{fig:unsurxilnuhf} as a function
of $\ln{U}$, for different strengths of the disorder $W$. The
inset presents $\log{D(M)}$ for $U=1, W=2$. The $U=0$ limit shows
that $\xi$ decreases when the disorder increases as for the
Anderson localization. For moderate interactions, an enhancement
of the localization length $\xi$ when compared to its
non-interacting value is observed for all disorder strengths $W$.
The values for $W=2$ are very high, even larger than the sizes of
the studied systems. In such a case, eq. (\ref{eq:loclexpression})
is not valid anymore. Yet, it is possible to conclude to a strong
and significant increase of $\xi$. The increase factors $\xi_{{\rm
max}}/\xi(U=0)$ reaches 2.2, 3.4 and 4.5, for $W=2,5$ and $9$,
respectively. They are significant and confirm the delocalizing
effect. The maximum of $\xi(U,W)$ belongs to the interval
$[\frac{W}{2}\pm t]$ and decreases as $W$ increases. This strong
enhancement suggests that the wave functions are less localized
and that the electrons might move more easily on several sites.
Moreover, the stiffness depends on the localization length
according to the equation:
$D(M)=(-1)^{M/2+1}M^{1/2}D(U,W)\exp{-M/\xi(U,W)}$
\cite{stafford92}. Hence, when $\xi$ increases, $D$ also increases
and the persistent currents will consequently increase even for
large finite rings of circumference $M$. Another weak increase for
the localization length had been reported for spinless fermions
but only for a strong degree of disorder \cite{rodolfo01}. In the
case of strong interactions, the localization length $\xi$
decreases with increasing $U$, as a Mott insulator. As for the
stiffness $D$ that was presented in Sec. \ref{sec:halffilling}
close to the strong interaction limits, the disorder phenomenon
has the unexpected effect of increasing the localization length.
This phenomena that is observed for systems of $M=20$ sites can be
extended to bigger systems.

\begin{figure}[h!]
    \begin{center}
        \includegraphics[width=0.4\textwidth]{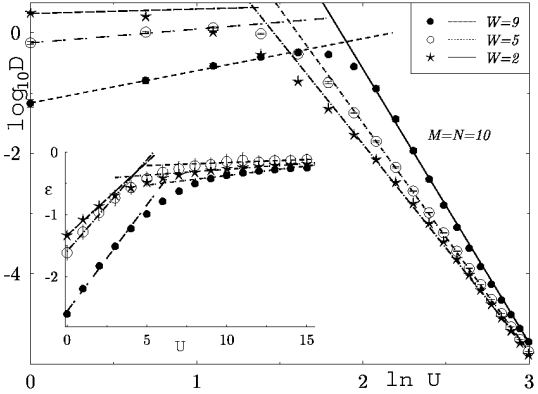}
    \end{center}
    \caption{$\log{D}$ ($\epsilon$) is plotted as a function of $\ln{U}$ ($U$) for half-filled systems
for $W=2,5,9$. The dotted lines are the  fitting for both regimes.}
    \label{fig:logulogD1010}
\end{figure}

In order to illustrate both competing regimes, one should fit the
inverse of the localization length. Fig. \ref{fig:unsurxilnuhf}
shows that $\xi^{-1}$ depends on $\ln{U/t}$ for all disorder
strength.
This dependence is expected for very strong interactions and $W=0$
\cite{stafford92}. However, for $W \ne 0$ and for weak
interactions, the inverse of the localization length behaves like
$\ln{U^{-|a|}}$ as a correlation length whereas for strong
interactions it behaves following $\ln{U^{+|b|}}$. For example
when $W=2$, $\xi^{-1}$ behaves as $-1.0+0.8\ln{U}$ for weak
interactions and then, as $0.022-0.001\ln{U}$ for strong
interactions, in Fig. \ref{fig:unsurxilnuhf}. In Fig.
\ref{fig:logulogD1010}, $\log{D}$ is represented as a function of
$\ln{U}$ and two different regimes also clearly appear. First $D$
increases up to $U^{+|a^{\prime}|}$ and then decreases up to
$U^{+|b^{\prime}|}$. This is true for all the systems that we
considered as well as for all non-zero disorders.
The energy density $\epsilon=\partial E/\partial M$ was
represented as a function of the interaction $U$, for different
disorder strengths $W=2,5,9$ (see inset of Fig.
\ref{fig:logulogD1010}). At half-filling, when $U$ is strong, the
particles tend to stay isolated within their site and the system
becomes rigid, as previously mentioned in Sec.
\ref{sec:halffilling}. The energies converge towards zero, and the
density of energy increases as $-\exp{-CU}\sim CU$. For very
strong interactions i.e. when there is exactly one particle per
site, there is little displacement and the energies correspond to
the total sum of the on-site potentials. As the systems are
half-filled and the potentials are random, the sum tends to
converge to zero for all degrees of disorder. The energy density
tends to saturate at zero. Thus, one fits $\epsilon$ for weak and
strong interactions with straight lines proportional to $U$.

Figures \ref{fig:unsurxilnuhf} and \ref{fig:logulogD1010} present
the two competing regimes and reveal a very sharp  crossover, as
the fittings joining at $U_{\mathrm{cross}}$ covers almost the
totally of the symbols used in the different curves. Here, we
illustrate two exclusion limits. For $U=0$, the electrons with
opposite spin remain on the same site (Pauli principle) while for
infinite interactions, the electrons avoid being on the same site
(the exclusion principle). For strong interactions, the mobility
of the electrons is weak since they are trapped alone within their
site and hence, the movement is frozen. When the interaction
reduces, electrons can jump on other sites and the crossover can
then occur. For weak interactions, the electrons can move easily
on multiple sites and the movement is dominated by quantum
fluctuations. The second derivative of the quantities $\xi^{-1}$,
$\log{D}$ and $\epsilon$ illustrate this crossover. For weak
interactions, $\partial^{2}\epsilon/\partial U^{2}$ behaves as
$-C^{2}\exp{-CU}$ while for strong $U$, it is a constant. Between
these limits, a peak exists and symbolizes the crossover. A study
of the shift for this peak as a function of the degree of disorder
should be considered to describe the corresponding crossover as a
function of the interaction, $U_{\mathrm{cross}}$ as shown in Fig.
\ref{fig:ucrossaway}.

\subsection{\label{ssec:loclengthaway} Localization length off half-filling}

The study of the localization length is now considered for off
half-filling. One considers three systems with a similar number of
particles $N=8$, but with different size $M=9,10,11$. Some of
these systems were presented in Sec. \ref{sec:awayhalffilling}.

\begin{figure}[h!]
    \begin{center}
        \includegraphics*[width=0.4\textwidth]{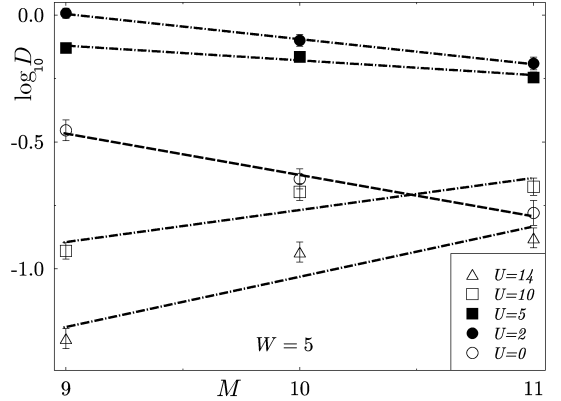}
    \end{center}
    \caption{$\log{D}$ is plotted as a function of $M=9,10,11$ for a degree of disorder $W=5$,
    for $N=8$ particles. The slope of $\log{D}$ changes sign as $U$ increases.}
    \label{fig:linesaway}
\end{figure}

\begin{figure}[h!]
    \begin{center}
        \includegraphics[width=0.4\textwidth]{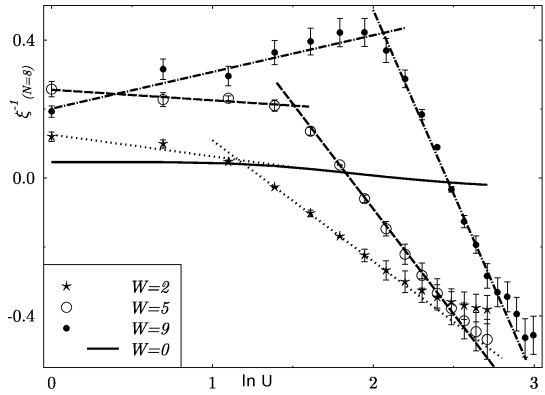}
    \end{center}
    \caption{$\xi^{-1}$ is plotted as a function of $\ln{U}$ for $W=0,2,5,9$, for $N=8$ particles.
    The straight lines correspond to the fits of both regimes. A crossover is here identified.}
    \label{fig:unsurxilnun8}
\end{figure}

Fig. \ref{fig:linesaway} illustrates $\log{D(M)}$ for different
degrees of interaction, with $W=5$. As in Fig.
\ref{fig:dwfixaway}, results reveal that for $U>W$, $\log{D(M)}$
increases as $M$ increases. The slope is thus positive. This leads
to a change in the sign of the localization length off
half-filling which is in agreement with the comments made
following eq. (\ref{eq:loclexpression}).
One can assume that approximately straight lines are present even
if only three points are here considered. At the point where the
curves $\log{D(U,M)}$ cross (see Fig. \ref{fig:dwfixaway}), the
localization length strongly increases and reaches a high positive
value. After the crossing, it reaches an important negative value
before increasing once more. Figure \ref{fig:unsurxilnun8}
illustrates this behavior for $W=0,2,5,9$, considering the inverse
of the localization length $\xi^{-1}$. When $U=W=0$, the
localization length starts at a very high value and decreases when
$U$ increases. When $U=0$, $\xi$ decreases as the degree of the
disorder increases as it is case for the Anderson insulator
\cite{lee85kramer93}. For $W=0$, $\xi$ decreases, then, a change
in sign occurs and $\xi$ increases up to a disorder-dependent
negative value. For all disorders $W \ne 0$ and moderate values of
$U$, $\xi$ increases strongly compared to its non-interacting
value. If one considers the positive values of $\xi$, it is
noticeable that $\xi(0)>\xi(2)>\xi(5)>\xi(9)$ (except for $U=2$,
due to numerical errors). If the highest positive value is not
considered, the ratios $\xi_{\mathrm{max}}/\xi(0)$ reach 1.5,
2.75, and 4 for $W=2,5$ and $9$, respectively. These ratios are
close to but smaller than those observed at half-filling. After
the change in sign, the localization length increases up to a
negative value $\xi_{\mathrm{neg}}$ that depends on the degree of
disorder. When the disorder increases, the localization length
decreases as expected for off half-filling (Sec.
\ref{sec:awayhalffilling}).

\begin{figure}[h!]
    \begin{center}
        \includegraphics[width=0.4\textwidth]{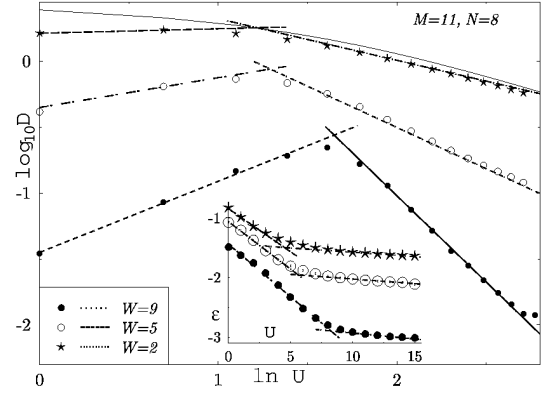}
    \end{center}
    \caption{$\log{D}$ ($\epsilon$) is plotted as a function of $\ln{U}$ ($U$) for $N=8$.
    The dotted lines correspond to the fitting of both studied regimes.}
    \label{fig:logulogD118}
\end{figure}

In order to study both regimes (Pauli and exclusion limits) as was
in Sec. \ref{ssec:loclengthhf} one fits $\xi^{-1}$ as a function
of $\ln{U}$ for weak and strong interactions in order to obtain
the $U_{\mathrm{cross}}$. This is presented in Fig.
\ref{fig:unsurxilnuhf}.
For positive values, the inverse of the localization length
$\xi^{-1}$ behaves as $\ln{U^{-|a_{2}|}}$ and for strong
interactions, the localization length behaves as
$\ln{U^{-|b_{2}|}}$ because of the change in sign. In Fig.
\ref{fig:logulogD118}, the curves plotting $\log{D}$ as a function
of $\ln{U}$, for various degrees of disorders, are represented and
it is possible to fit them with lines proportional to $\ln{U}$.
They present behavior similar to that observed for half-filling
behaviors.
The energy density $\epsilon = \partial E/\partial N$ is
represented as a function of the interaction $U$, in the inset of
Fig. \ref{fig:logulogD118}. For weak interactions, the electrons
are placed by pairs (with opposite spin directions) within the
same site. In addition, the corresponding energies are summed over
$N/2$ sites. When the interactions become stronger, the energies
increase as it would be the case for sums over less negative
potentials. The energies of the ground-state when the number of
sites increases, decrease strongly. This leads to a decrease in
the energy density as expressed by $\exp{-CU}\sim -CU$. For very
strong interactions, the ground-state energies are summed over
on-site potentials. As there are empty sites, the energy density
tends to saturate at a specific negative value that increases with
the strength of the disorder.

The quantities $\epsilon$, $D$ and $\xi$ highlight the presence of
a crossover as mentioned in Sec. \ref{ssec:loclengthhf} and gives
the means to determine the existence of a peak between the two
different regimes. The position of the peak is given by the value
$U_{\mathrm{cross}}$, when the curves fitting both regimes cross.
\begin{figure}[h!]
    \begin{center}
        \includegraphics[width=0.4\textwidth]{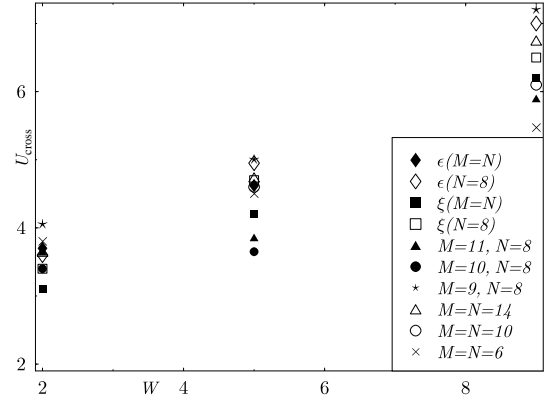}
    \end{center}
    \caption{$U_{\mathrm{cross}}$ as a function of the disorder $W$ for systems at and off half-filling.}
    \label{fig:ucrossaway}
\end{figure}
In Fig. \ref{fig:ucrossaway}, the movement of the peaks is a
function of the strength of the disorder $W$, for systems at and
off half-filling. When $W$ increases, the interaction must be
stronger in order to reach the second regime where the particles
will be localized alone within their site. The peaks and hence,
the crossovers occur quickly around the same value of the
interaction $U$, for a given disorder $W$. This is true for all
systems. The values of $U_{\mathrm{cross}}$ are very close but
more scattered for off half-filling. The competition between the
two regimes that are represented by the Anderson and the Mott
insulators is very important in order to gain a better
understanding of the two dimensional metal-insulator transition,
which is at the heart of the problems in mesoscopic physics
\cite{kravchenko94}. With disorder only, no metal-insulator
transition is found \cite{lee85kramer93}. When including these
interactions, it is possible to determine two different regimes in
one dimension. This may explain the metal-insulator transition in
two dimension situation.

\section{\label{sec:conclusion} Conclusion}

We investigated the effect of electron-electron on-site
interactions on the persistent currents found in disordered
one-dimensional rings with arbitrary fillings. We took into
account the spin and considered the stiffness as a measure of the
magnitude of the persistent currents. Our numerical simulations
indicate a strong increase in the stiffness with respect to the
non-interacting case observed for moderate interactions and
arbitrary fillings. This strong enhancement, which is also
observed at the level of disordered realization, can be attributed
to the mechanism of both competing forces, i.e. disorder and
interaction. A simple computation comparing the weakly interacting
state systems in the presence of disorder shows an increase in the
electrons mobility within the ring. The stiffness decreases for
stronger interactions according to a power law that is different
at and off half-filling. The analytical approaches helped us to
determine the power laws for both cases. An unexpected behavior
was also observed at the strong interaction limit, specifically at
half-filling: the disorder rendered the stiffness stronger. Such
an effect can be explained by the disorder-induced reduction in
the energy gap between the Mott insulator ground-state and the
excited states. By studying the localization length, we here
further observed that the interactions increase the persistent
currents even for a large finite size of rings at half-filling.
The localization length, which is defined phenomenologically,
changes sign off half-filling. The behavior of the energy density,
the stiffness as well as the localization length suggest a
crossover between weak and strong interactions. Based on these
results, we can conclude that the spin plays a significant role
and influences the increase in magnitude of persistent currents.

\appendix

\section{\label{sec:perturbationexpansion} Perturbation expansions}

\subsection{\label{sec:analyticshf} Half-filling}

We give here the details for the analytical approach in Sec.
\ref{sec:halffilling}. The ground-state is given by eq.
(\ref{eq:exactGShalf}) for both periodic and antiperiodic
conditions \cite{lieb62,lieb93}. The aim here was to define
sequences (the set of hops defined in Sec. \ref{sec:halffilling})
that may lead to discrepancies within the corrections brought to
the energy level.

The terms presented in the numerator of $E^{(n)}$ are equal to
$-t$ but the signs of the permutations of the electrons, with spin
up or down, could be negative as the hopping element changed the
electrons' order. For periodic boundary conditions, the numerator
can be written as $N_{p}=(-t)^n \mathrm{sign}(P_{\uparrow})
\mathrm{sign}(P_{\downarrow})$ where we define
$P_{{\uparrow}({\downarrow})}({\bf A})$ $\uparrow$($\downarrow$)
as the permutation of the positions for up (or down) electrons
around the ring, which is a resultant from the defined sequence
${\bf A}$. Since the flux appears only within the kinetic part
$H_K$ of the Hamiltonian, a unique change only appears when a
particle crosses the border for antiperiodic boundary conditions.
This introduces a sign $(-1)$ in the numerator. Thus, the
numerator for antiperiodic boundary conditions is equal to :
$N_{ap}=(-1)^{h_b}~N_{p}$ where $h_b$ is the number of hops across
the border $1~\longleftrightarrow~M$ contained in ${\bf A}$.
For orders lower than $M$, the $n^{\rm th}$ order energy is a
constant in $\Phi$ that does not depend on the boundary conditions
and the phase sensitivity is zero. Indeed, an even number of hops
is needed to recover one of the eigenstates and in this case, the
border is crossed twice. To obtain a difference, the particles
must circulate around the ring and the lowest contributions are of
order $M$; in such case, a particle crosses the boundary condition
only once. Our work gave us the means to define different
sequences connecting two basis-states, $|\Psi_{\beta} \rangle$ and
$|\Psi_{\beta}^{\prime} \rangle$, and to obtain eq.
(\ref{eq:Wexpansionfinal}). These sequences have previously been
published \cite{gambetti02}.

When disorder is reintroduced, the perturbation theory remains
second order, with an exchange energy that is now equal to
$\frac{2t^2}{U(1-\frac{W^2}{U^2}(v_i-v_{i-1}))^2}$. The matrix
elements of the Hamiltonian are under such conditions
even-functions of $\frac{W}{U}$, so are the weights $f_{\beta}$.
They read: $f_{\beta}\approx
f_{\beta}^{(0)}+f_{\beta}^{(2)}(\frac{W}{U})^2$. In the following
development, we maintained $f_{\beta}^{(0)}$ only and $f_{\beta}
\sim f_{\beta}^{(0)}$ was approximated. In order to study the
effect of disorder on the phase sensitivity for strong
interactions ($U\gg W$), we expand the denominator under powers of
$W/U$. When second order is reached, this yields :
\begin{eqnarray}
\label{eq:Wexpansion}
\Delta E^{(M)}\approx\frac{
  (-1)^{N/2}4t^{M}}{U^{M-1}}\sum_{\beta,\beta'}\sum\limits_{{\bf A}_{\rm f}^{(\beta,\beta')}}
\frac{f_\beta f_{\beta'}}{\prod_{l}g_{\gamma_l}} \nonumber\\
\left(1-\frac{W}{U}\sum_{l}\frac{d_{\gamma_l}}{g_{\gamma_l}}
+\frac{W^2}{U^2} \left(\sum_l\frac{d_{\gamma_l}^2}{g_{\gamma_l}^2}
+\sum_{l<m}\frac{d_{\gamma_l}d_{\gamma_m}}
                {g_{\gamma_l}g_{\gamma_m}}\right)\right).
\end{eqnarray}
The dominating term of (\ref{eq:Wexpansion}) corresponds to the
power law found in section \ref{sec:halffilling}. We now consider
the term of the first order in $W/U$ and this first-order
correction for $W/U$ vanishes because of the particle-hole
symmetry. Indeed, if one characterizes a given sequence of
hoppings by the positions of the doubly-occupied and the empty
sites, one can always construct moving the particle along the ring
a second sequence by exchanging the places of the doubly-occupied
and the empty sites. These two sequences will possess an equal
number of doubly-occupied sites $g_{\gamma_l}$ and their
coefficients $d_{\gamma_l}$ will have opposite signs. This first
order term vanishes when the sum over all the sequences is taken.
Thus, the second order term of (\ref{eq:Wexpansion}) determines
the disorder dependence of the persistent current for $U \gg W$.
The first term is of order $1$ and is always positive. Hence, for
each sequence the sum over positive quantities of order one only
are taken and it is possible to estimate: $\sum_{l}
(\frac{d_{\gamma_{l}}}{\epsilon_{\gamma_{l}}})^2 \lesssim M$. By
summing over all the sequences, we obtain a positive term of order
$N_{S}M$. The second term is also of order $1$ but has a random
sign. Taking the sum over all sequences, it is evaluate to almost
$\frac{\pm \sqrt{N_{S}M}}{\sqrt{2}}$. Since the number of
sequences is very high, one obtains $\sqrt{N_{S}} \ll N_{S}$. This
results in eq. (\ref{eq:Wexpansionfinal}). It is concluded here
that the phase sensitivity possesses positive corrections of order
$(\frac{W}{U})^2$, so does the stiffness $D$.

\subsection{\label{sec:analyticsaway} Off half-filling}

We give here the details of the calculations for the perturbation
expansion that was presented in section \ref{sec:awayhalffilling}.
The ground-state is given by eq. (\ref{eq:GSAHFW=0}).

\begin{figure}[h!]
    \begin{center}
\includegraphics*[width=0.3\textwidth,height=3.5cm]{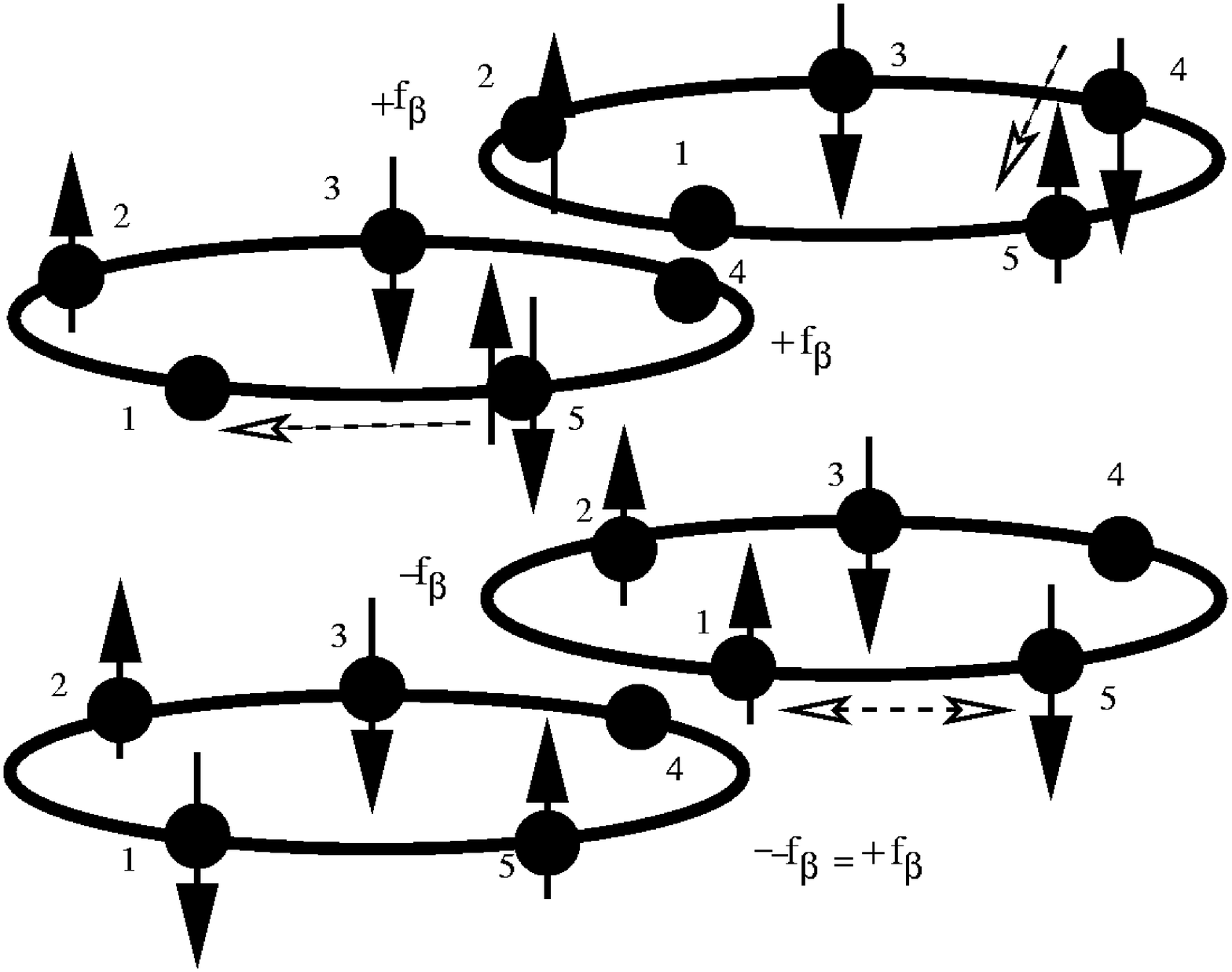}
    \end{center}
    \caption{Example of a ``double jump'' sequence for $M=5$ and $N=4$. The third and fourth drawing illustrates
an "exchange" sequence.}
    \label{fig:doublejump}
\end{figure}

The matrix element of second order contains intermediate states
with exactly one doubly-occupied site, $|\psi_{\gamma} \rangle$'s,
that connect $|\psi_{\beta}^{{\cal C}_i} \rangle$ to
$|\psi_{\beta}^{{\cal C}_j} \rangle$, as the second jump leads to
$|\Psi_{\beta} \rangle$. We seek to develop sequences of two jumps
that will lead to different terms ${\cal H}^{(i,j)}$, according to
the defined boundary conditions. Then, crossing border sequences
are considered. The first sequences are the exchange sequences,
where two electrons exchange places with one another, across the
border, as illustrated in Fig. \ref{fig:doublejump}. Under such
conditions and for both periodic and antiperiodic boundary
conditions, the factor $F=\langle \psi_\beta^{{\cal
C}_i}|H_{K}|\psi_{\gamma}\rangle
\langle\psi_{\gamma}|H_{K}|\psi_{\beta}^{{\cal C}_j}\rangle$ is
equal to $t^{2}$. More specifically, under periodic boundary
conditions, weights have equal signs, whereas under antiperiodic
boundary conditions, weights have opposite signs because of the
Marshall's rules \cite{weng91,marshall55}. A second type of
sequence, called the double jump is illustrated in Fig.
\ref{fig:doublejump}, and brings different terms ${\cal
H}^{(i,j)}$ to the Hamiltonians. Under periodic and antiperiodic
boundary conditions, the factor $F$ is equal to $t^{2}$ and
$-t^{2}$, respectively. In addition, under periodic boundary
conditions, weights maintain equal signs. However, under
antiperiodic boundary conditions, the double jump is equivalent to
two hops for a given particle i.e. equivalent to a change in sign
and an exchange. This is also illustrated in Fig.
\ref{fig:doublejump}. Overall, this second type of sequence gives
an end product with an equal sign for the weights of both
$f_{\beta}^{{\cal C}_i}$ and $f_{\beta}^{{\cal C}_j}$.
This leads to a contribution of a second order phase-sensitivity.
The factorization of $-t^{2}/U$ for all the different terms of
both Hamiltonians leads to eq. (\ref{eq:deltae2away}).

When disorder is introduced, the ground-state is a superposition
of the basis-states within each subspace ${\cal C}^{R}_{i}$ with
weights $f_{\beta}$ chosen to be real. At this point, one {\it
assumes} that the Marshall rule is still valid in presence of
disorder \cite{marshall55,perronfrobenius}. Due to signs of the
weights, the only possible sequences that could lead to opposite
values for periodic and antiperiodic boundary conditions are the
"exchange" sequences. Now, using sums over two-hops sequences
${\bf A}^{(\beta,\beta')}$, the second-order phase sensitivity
reads:
\begin{eqnarray}
\label{eq:DeltaE2awaywithdisorder} \Delta E^{(2)}=
\frac{-2t^{2}}{U}.\frac{\kappa^{\prime}}
{(1-\frac{W^{2}}{U^{2}}d^{2}_{\gamma})}
\end{eqnarray}
The expansion of eq. (\ref{eq:DeltaE2awaywithdisorder}) leads to
the eq. (\ref{devorder2}).

\subsection{\label{sec:toymodel2on3} Toy model}

The toy model is presented here to help explain the stiffness
behavior that was presented in Sec. \ref{sec:awayhalffilling}, for
general systems of $N$ electrons on $M$ sites. Here, we consider a
simplified model of two electrons for three sites. Such a model
was studied by H. Tasaki and provided the means to verify parts of
our results \cite{tasaki95}. This small system is described by the
Hubbard-Anderson Hamiltonian of eq. (\ref{eq:generalHamiltonian}),
when restricted to three sites. In this case, one considers one
empty site only \cite{tasaki95,lieb93}, as Sec.
\ref{sec:awayhalffilling}.

We first use numerical manipulations in order to compute the
many-body ground-state energies and to deduce the phase
sensitivity. The basis-states are given by eq. (\ref{basicstate}).
Thus, diagonalization of the matrices is necessary to obtain the
eigenstates for all values of the parameters $t$, $U$ and $v_i$.
The case without disorder is now considered. Under periodic
boundary conditions, the energy level is described by
$E(0)=\frac{1}{2}U-t-\frac{1}{2}\sqrt{U^2+4Ut+36t^2}$. As far as
the antiperiodic conditions are concerned, the minor eigenvalue is
equal to $E(\pi)=-2t$. Now it is possible to calculate, using eq.
(\ref{eq:phasesensitivity}), the contributions to the phase
sensitivity. This would lead to:
\begin{eqnarray}
    \label{eq:edevinteraction}
\Delta E =&E(0)-E(\pi) \nonumber\\
&=t+\frac{1}{2}U-\frac{1}{2}\sqrt{(U^2+4Ut+36t^2)}
\end{eqnarray}
For very strong interactions, an expansion in $t/U$ can be
performed and the phase sensitivity is then given by:
\begin{eqnarray}
\label{eq:phsens2on3strongUnum}
\Delta E \approx -\frac{8t^2}{U}
\end{eqnarray}
This quantity behaves as $1/U$ and decreases when $U$ increases.
For periodic boundary conditions, the ground-state of the system
is a superposition of the states $|\Psi_{\beta} \rangle$ and
$|\Psi_{\gamma} \rangle$. All weights are positive. Indeed, the
unnormalized weights for the states $|\Psi_{\beta} \rangle$ are
all equal to one whereas those for the states $|\Psi_{\gamma}
\rangle$ are equal to $f^{*}_{\gamma}$. One assumes ${\cal
N}^{2}=3(f^{*}_{\gamma})^{2}+6$. The components of the
$|\Psi_{\gamma}\rangle$'s are written as follows:
\begin{eqnarray}
    \label{eq:weightsdoublyoccupied}
    f_{\gamma}=\frac{f^{*}_{\gamma}}{\cal N}=-\frac{1}{2}\frac{\frac{1}{2}U+t-\frac{1}{2}\sqrt{U^2+4tU+36t^2}}{t {\cal
    N}}
\end{eqnarray}
The development of these weights $f_{\gamma}$ for very large $U$
is such that (see eq. (\ref{eq:weightsdevinteraction})):
\begin{eqnarray}
    \label{eq:weightsdevinteraction}
    f_{\gamma} \approx \frac{8t}{2U \cal N}.
\end{eqnarray}
The weights of the $|\Psi_{\gamma} \rangle$'s decrease down to
zero whereas the weights of the states $|\Psi_{\beta} \rangle$ are
equal to
\begin{eqnarray}
    \label{eq:fbetatoymodel}
    f_{\beta} = \frac{1}{\cal N} \approx \frac{1}{\sqrt{6}}.
\end{eqnarray}
We consider that for the ground-state and within the limits of
very strong interactions, one can neglect these $|\Psi_{\gamma}
\rangle$'s. Under antiperiodic boundary conditions, the
corresponding eigenvector is the superposition of the
$|\Psi_{\beta} \rangle$'s with alternated signs.

Second, it is possible to calculate the corrections to bring to
the energy level using the perturbation theory in order to recover
the values of the phase sensitivity given by eq.
(\ref{eq:phsens2on3strongUnum}). The basis-states remain the $\cal
S$-states. The first state has an $\uparrow$-electron on the first
site and a $\downarrow$-electron on the second site, the last site
being empty in all cases. The other states are generated by one
hop processes. This set of six states constitutes a
"super-lattice" \cite{nagaoka66} and is a good illustration of the
connectivity condition \cite{tasaki95}. The basic structure of the
ground-state is determined following the Perron-Frobenius sign
convention \cite{tasaki95,lieb93,perronfrobenius}. The
ground-states are in fact the superposition of the six states,
denoted $|1 \rangle \cdots |6 \rangle$, with signs respecting the
Marshall's rule \cite{marshall55} as indicated in the following:
\begin{eqnarray}
    \label{eq:exempleetat2sur3}
 &|\Psi_{0,\Phi=0(\Phi=\pi)} \rangle = \frac{1}{\sqrt{6}} \nonumber\\
 & \left( |1 \rangle + |2 \rangle +|3 \rangle +\left( -\right)|4 \rangle +\left( -\right)|5 \rangle
 +\left( -\right)|6 \rangle\right)
\end{eqnarray}
We can now calculate the periodic and antiperiodic many-body
first-order energies levels. Under periodic boundary conditions,
each $|\psi_{\beta} \rangle$ is coupled to two other states by a
single jump, and all terms are positive $E(0)=\frac{1}{6}(-t)
\times 2 \times 6=-2t$. Under antiperiodic boundary conditions,
the negative sign that is induced by the flux, is compensated by
the negative sign of the weights each time the electron crosses
the border line. One gets $E(\pi)=-2t$. This result shows that the
phase sensitivity cancels at first order, as explained in Sec.
\ref{sec:awayhalffilling}. Similar eigenvalues are obtained
through the numerical diagonalizing of the matrices that are
formed with the basis-states $|1\rangle \cdots |6\rangle$. The
calculations must be increased to the second order in order to
recover a contribution. Hence, one must consider two-hops
sequences across the border. The first jump leads to a
doubly-occupied site whereas the second jump leads to one of the
many basis-states. Each state $|i \rangle~(i=1, \ldots, 6)$ is
connected to two states $|\Psi_{\gamma} \rangle$. Each of these
$|\Psi_{\gamma} \rangle$'s can lead to 4 different $|\Psi_{\beta}
\rangle$'s. Under periodic boundary conditions, since the weights
are equal and posses similar signs, the contribution is
$E^{(2)}(0)=\frac{1}{6U}[6 \times 2 \times 4 \times
(-t^{2})]=-8t^2/U$. In the case of antiperiodic conditions, the
sum over these terms cancels out. Indeed, if we now consider the
state $|1 \rangle$, half of the terms contain a negative sign and
the other half a positive one and hence, the contribution is zero.
A similar comment can be given to other states, thus, the phase
sensitivity is equal to $-8t^2/U$ as was obtained in eq.
(\ref{eq:phsens2on3strongUnum}). This confirms the validity of the
perturbation theory within the limit for very strong interactions.

In this toy model one introduces a disorder $v$ on the first site,
when considering the limits for very strong interactions. The
$6*6$ matrix can be diagonalized with the disorder $v$ placed upon
the first site. This yields :
\begin{eqnarray}
&E_{{\rm periodic}}= E_{{\rm antiperiodic}} = \frac{(E(v)+E(-v))}{2}\nonumber\\
& = \frac{1}{2}t -
\frac{1}{4}\sqrt{v^2-2tv+9t^2}-\frac{1}{4}\sqrt{v^2+2tv+9t^2}.
\label{eq:edistoymodel}
\end{eqnarray}
Even in the presence of disorder, the first-order phase
sensitivity is equal to zero. Hence, for strong disorders,
particles occupy the disordered states associated to the following
weight:
\begin{eqnarray}
f_{\beta}= & \frac{\frac{1}{2}v - \frac{1}{2}t -
\frac{1}{2}\sqrt{v^2-2tv+9t^2}}{t {\cal N}^{\prime}}
 & \approx_{v\to \infty} \frac{9t^2}{4v {\cal N}^{\prime}}
\label{eq:fbetadistoymodel}
\end{eqnarray}
where $({\cal N}^{\prime})^{2}=\sum_{\beta}(f^{*}_{\beta})^{2}$.
The $f_{\beta}(W\to \infty)$ decrease when $W$ increases. This
gives the means to avoid the sites with highest potentials in Sec.
\ref{sec:awayhalffilling}.

\section{\label{sec:infUweightsGS} Very strong interactions limit}

\begin{figure}[h!]
\includegraphics*[width=0.4\textwidth]{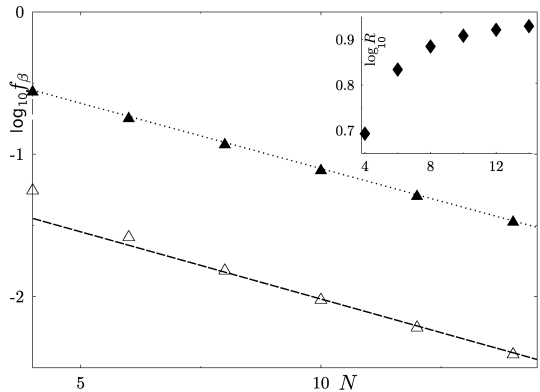}
\caption{Logarithm of the most significant components for the
ground-state of half-filled systems ($N=M$) (in the inset is
presented the logarithm of their ratio
$R=f_{\mathrm{alt}}/f_{\mathrm{max2}}$) for infinite interactions
as a function of the system size, $M$. }
    \label{fig:logfbeta}
\end{figure}

This part concerns the infinite interaction case. The ground-state
components at half-filling were considered. A program was
developed in order to calculate the weights of the eigenvectors
when two particles can not remain together upon the same site ($U
\to \infty$). We computed the weights of each component for the
ground-state as a function of the ring size $M$, for $M \in
[2,14]$. The most significant weights, $f_{\mathrm{alt}}$, are
those of the states with alternated spins; the second most
important ones, $f_{\mathrm{max2}}$, correspond to states where
two electrons have exchanged their positions when compared to the
states with alternated spins. We have represented the logarithm of
these $f_{\mathrm{alt}}$ and $f_{\mathrm{max2}}$ weights as a
function of the size of the system $M$ (see Fig.
\ref{fig:logfbeta}). It was established that the weights $f(M)$
can be fitted by an exponential. This result gave the means to
extrapolate what would be obtained in the case of large $N$: we
speculate that for big systems, the alternated spin configurations
will be the most significant contributor to the phase sensitivity
parameter. Moreover, as the ratio $R$ saturates and as soon as $M
\ge 12$, one may conclude that the corrections brought to the
phase sensitivity will essentially depend on both
$f_{\mathrm{alt}}$ and $f_{\mathrm{max2}}$ weights. Here, the
particles move easily around the ring, especially for the states
with alternated spins. Consequently, an important current may be
induced.

\section{\label{sec:oddnb} Odd number of electrons}

\begin{figure}[h!]
    \begin{center}
        \includegraphics*[width=0.4\textwidth]{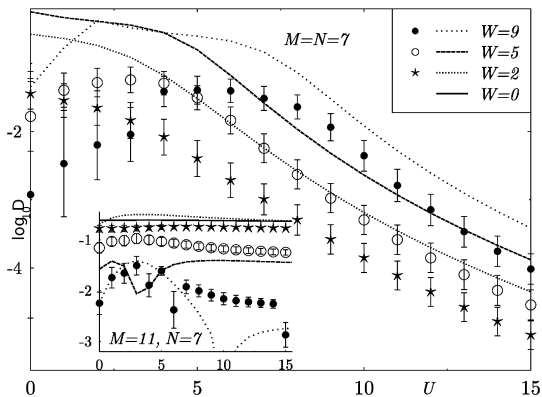}
    \end{center}
    \caption{$\log{D}$ as a function of the interaction $U$ for $W=(0,)2,5,9$ for $M=N=7$ (for $M=11$, $N=7$ in the inset).
    For intermediate levels of interactions, a curve increase is observed. Specifically, for the sample and $W=9$,
    a minimum occurs at $\log{D}=-4.2$.}
    \label{fig:oddnb}
\end{figure}

Here, we consider the rings filled with an odd number of
electrons. The behavior is presented in Fig. \ref{fig:oddnb}, for
two systems $M=11$, $N=7$ and $M=N=7$.
For a disorder equal to zero, the phase sensitivity is equal to
zero for the half-filled case, $M=N=7$. The results show for the
sample, an absence of Anderson localization for $U=0$. Moreover
the sign of the phase sensitivity depends on the on-site disorder
potentials and is random. This renders the averaging process
difficult. Mean averages must be performed on $\Delta E$, and the
fluctuations are in such a case much more significant than for
systems with an even number of electrons. This is shown in Fig.
\ref{fig:oddnb}. The averaged stiffness terms
$\langle\log{D}\rangle$ respect Anderson localization for $U=0$
for both systems. For moderate values of the interactions and for
the disorders $W=2,5,9$, an increase of $\log{D}$ is observed for
the mean calculated averages and some various studied samples
(except here for $W=2,5$ at half-filling). But in all cases,
$\log{D}$ is smaller than that observed for an even number of
electrons. Nevertheless $\log{D}$ can reach an order of magnitude
($D_{\rm max}/D(U=0)\sim 34$), at half-filling ($M=N=7$) and for
$W=9$. For stronger interactions, the stiffness parameter
decreases with $U$. At half-filling, the disorder increases the
magnitude of the persistent current within the very strong
interaction limits as what was observed for the case of even
numbers of electrons. Finally, a perturbative approach is not
possible as those theorems are applied uniquely for the case of
even numbers of both sites and electrons.

\begin{acknowledgments}

Supercomputer time was provided by CINES (project gem2381) on the
SGI 3800. I am grateful to Janos Polonyi and Thierry Giamarchi for
their enthusiastic discussions and for reading the manuscript. I
also thank Jean Richert for reading the manuscript and Daniel
Cabra for useful comments on the manuscript.

\end{acknowledgments}


\end{document}